\documentclass[12pt]{article}
\usepackage[margin=0.95 in]{geometry}
\usepackage{amsmath}
\usepackage{amssymb,amsfonts}
\usepackage[all]{xy}
\usepackage{graphicx}
\usepackage[utf8x]{inputenc}
\usepackage{amsmath}
\usepackage{amssymb}
\usepackage{float}
\usepackage{array}
\usepackage{tikz}
\usepackage{mathtools}
\usepackage{mathrsfs} 
\usepackage{hyperref}  
\usepackage{cite}
\numberwithin{equation}{section}
\numberwithin{equation}{section}
\numberwithin{table}{section}
\newcommand{\Tr}{\mathrm{Tr}\,}
\newcommand{\p}{\partial}
\newcommand{\ext}{\text{ext}}
\newcommand{\op}{\text{op}}
\newcommand{\cl}{\text{cl}}
\begin{document}

\thispagestyle{empty}

\vspace*{3cm}
{}

\noindent
{\LARGE \bf Topological Open/Closed String Dualities:}
\vskip .23cm
\noindent
{\LARGE \bf 
Matrix Models and Wave Functions }
\vskip .35cm
\noindent
\linethickness{.05cm}
\line(10,0){447}
\vskip 1.0cm
\noindent
\noindent
{\large \bf Sujay K. Ashok$^a$ and Jan Troost$^b$}\\

{\em 
\vspace{ 0.1cm}
\noindent
$^a$Institute of Mathematical Sciences, 
Homi Bhabha National Institute (HBNI),\\
IV Cross Road, C.~I.~T.~Campus, 
Taramani, Chennai, 600113  Tamil Nadu, India \\

\vspace{.2cm}
\noindent
$^b$Laboratoire de Physique 
de l'\'Ecole Normale Sup\'erieure \\
\hskip -.05cm
 CNRS,  
 PSL Research University and Sorbonne Universit\'es, Paris, France
}
\vskip 1.2cm

\vskip0cm

\noindent
{\sc Abstract: } {We sharpen the duality between open and closed topological string partition functions for topological gravity coupled to matter. The closed string partition function is a generalized Kontsevich matrix model in the large dimension limit. We integrate out off-diagonal degrees of freedom associated to one source eigenvalue, and  find an open/closed topological string partition function, thus proving open/closed duality. We match the resulting open partition function to the generating function of intersection numbers on moduli spaces of Riemann surfaces with boundaries and boundary insertions. Moreover, we connect our work to the literature on a wave function of the KP integrable hierarchy and clarify the role of the extended Virasoro generators that include all time variables as well as the coupling to the open string observable.  }

\vskip 1cm

\pagebreak

\newpage
\setcounter{tocdepth}{2}
\tableofcontents

\section{Introduction}

Two-dimensional gravity is an important tool in the study of quantum gravity. It  is a simpler analogue of four-dimensional gravity, and it arises in the diffeomorphism invariant theory living on a string theory world sheet.
Two-dimensional  quantum gravity has been solved to a large extent, in three ways: via double-scaled matrix models \cite{Brezin:1990rb,Douglas:1989ve,Gross:1989vs}, using conformal field theory techniques \cite{Knizhnik:1988ak,David:1988hj,Distler:1988jt} and in its topological formulation \cite{Witten:1988ze,Witten:1989ig,Kontsevich:1992ti}. The three approaches were proven to be equivalent in many instances. See e.g. \cite{Ginsparg:1993is,DiFrancesco:1993cyw} for reviews of the two-dimensional theory of quantum gravity.

While the topological formulation of two-dimensional gravity on Riemann surfaces without boundary was put on a firm footing a while back \cite{Witten:1988ze,Witten:1989ig}, the theory on Riemann surfaces with boundary was only recently understood rigorously \cite{Pandharipande:2014qya}. This gave rise to a flurry of mathematical activity which made earlier observations in the physics literature \cite{Dalley:1992br,Johnson:1993vk,McGreevy:2003kb,Gaiotto:2003yb,Maldacena:2004sn} precise \cite{Alexandrov:2014gfa,Buryak:2015eza,BCT1,Buryak:2018ypm}.
See \cite{Dijkgraaf:2018vnm} for a partial review peppered with physical insight and \cite{Muraki:2018rqv,Aleshkin:2018klz,Alexandrov:2019wfk} for more recent results.

Of particular interest to us here is the open/closed duality which was understood fairly well in the string theory literature \cite{Dalley:1992br,Johnson:1993vk,McGreevy:2003kb,Gaiotto:2003yb,Maldacena:2004sn}, and which was rigorously derived, including important additional details, in the more recent mathematical physics literature \cite{Alexandrov:2014gfa,Buryak:2015eza,BCT1,Buryak:2018ypm} in the case of pure two-dimensional gravity. This duality relates an open/closed string partition function to a purely closed string partition function; to be precise, the addition of a D-brane in topological string theory is transmuted into a shift of the background, a renormalization of the partition function, an operator insertion and an integral transform. While the conceptual framework goes back to \cite{McGreevy:2003kb,Gaiotto:2003yb,Maldacena:2004sn},  precise formulas were given more recently in \cite{Buryak:2015eza}. Only the latter allow to make direct contact with the rigorous analysis in algebraic geometry \cite{Pandharipande:2014qya,Buryak:2015eza,BCT1,Buryak:2018ypm}.

The main mathematical tool in the derivation of the precise form of open/closed string duality \cite{Buryak:2015eza} was the Kontsevich matrix model of pure two-dimensional gravity \cite{Kontsevich:1992ti}.
This matrix model depends only on the closed string sources, and can be viewed as a closed string matrix model in that sense.
The closed versus open/closed duality in the case of pure gravity was  derived in a particularly clear manner by integrating out off-diagonal degrees of freedom of a $(N+1) \times (N+1)$ matrix model to obtain a $N \times N$ matrix model depending on one extra eigenvalue which is the integral transform of an open string coupling (or D-brane modulus) \cite{Buryak:2015eza}.  In the large $N$ limit, this then gives rise to a duality between a closed string theory and an open/closed string theory. See also   \cite{Brezin:2011ka,Alexandrov:2014gfa} for related results.

In this paper, we find an alternative to   the original mathematical derivation that allows for a generalization  to the case of two-dimensional topological gravity coupled to matter. The open/closed duality in this case was understood to a degree in \cite{Hashimoto:2005bf}. We provide the mathematically precise relation between the open/closed and closed string partition function using the technique described above of integrating out off-diagonal degrees of freedom at finite $N$, and then taking the infinite $N$ limit. This will lead to considerable additional insight. Indeed, we naturally find a determinant insertion in the Kontsevich matrix model. We moreover find  that the closed string integrable hierarchy is intuitively extended to include the missing times. Importantly, by matching to the integrable system literature, we are able to obtain a precise algebro-geometric understanding of the resulting open/closed and extended partition functions. We also improve our understanding of the  Virasoro algebras that govern the generating functions.

The paper is structured as follows. In section \ref{integratingout}, we first present the result of partially integrating out off-diagonal matrix elements in the Kontsevich matrix model in an elementary fashion. We then interpret the resulting equations in terms of concepts in the integrable systems literature, as well as in string theory. We moreover build the bridge to the open/closed topological intersection numbers studied in algebraic geometry.
In section \ref{Virasoro} we provide a guide to the various Virasoro algebras that appear in the context of these integrable systems, and clarify how to interpret them. The knowledge gained is used to prove a crucial statement in section \ref{integratingout}. In section \ref{conclusions}, we  conclude with a summary and some open problems. In appendix \ref{FT} we provide properties of a generalized Fourier transform, in appendix \ref{AlternativeVirasoro} the result of conjugating Virasoro generators to a more familiar form, and  in appendix \ref{alternativepuregravity}, we review an alternative viewpoint on open/closed duality in the case of pure gravity.

\section{The Open/Closed String Duality with Matter}
\label{matter}
\label{integratingout}
\label{matrixintegral}
In this section, we first derive a central technical result in an elementary manner. 
We start from the generalized Kontsevich matrix model \cite{Kontsevich:1992ti,WittenAlgebraic,Adler:1992tj,Itzykson:1992ya} with matrix integration variables of dimension $(N+1) \times (N+1)$ and integrate out the off-diagonal matrix elements compared to $N \times N $ and $1 \times 1$ blocks around the diagonal. As we will discuss, the resulting equation can be interpreted in terms of standard concepts in integrable systems, and has interesting conceptual consequences in topological gravity and low-dimensional string theory.

\subsection{A Brief Overview}
%Orientation Briefing

The generalized Kontsevich model is a matrix generalization of the higher Airy function \cite{Kontsevich:1992ti}. 
When the matrix model integration variable tends to infinite size, it becomes a generating function for the $p$-spin intersection numbers on moduli spaces of Riemann surfaces \cite{Kontsevich:1992ti,WittenAlgebraic,Adler:1992tj,Itzykson:1992ya,Chiodo,Faber:2006gca,Brezin:2012uc}. Equivalently, it is the partition function of topological gravity coupled to matter of type $p$, or a closed string theory of gravity plus matter in dimension smaller than one \cite{Dijkgraaf:1990nc,Li:1990ke,Dijkgraaf:1990dj}. 

Our starting point is the study of the generalized Kontsevich model at finite $N$. In particular, we will
perform the integration over $2N$ off-diagonal degrees of freedom in the $(N+1) \times (N+1)$ matrix integration variables, and will be left with a $N \times N$ matrix model and an effective action for one more diagonal degree of freedom. In the large $N$ limit, both the $(N+1)\times(N+1)$ and the $N\times N$ dimensional model will correspond to closed string matrix models. The extra eigenvalue will eventually be  related  to a D-brane or open string modulus. The equation that results from the integration will thus allow for an interpretation in terms of open/closed string duality.
Equivalently, it allows for an interpretation of intersection numbers of Riemann surfaces with boundaries and bulk and boundary insertions, in terms of intersection numbers of Riemann surfaces with only bulk insertions.

%A number of sources of inspiration for our approach exist. 
Our approach has been inspired by a number of sources. Firstly, the idea of open/closed string duality can at least be traced back to the advent of D-branes in string theory \cite{Polchinski:1995mt}. The fact that it takes a particularly simple form in the case of pure topological gravity was understood in \cite{Gaiotto:2003yb}. These ideas were extended to the case of topological gravity with matter in \cite{Hashimoto:2005bf}.
At a technical level, these references have a  different approach. Secondly, the more rigorous reference \cite{Buryak:2015eza} follows almost the same technical route we described above to render the physical intuition in \cite{Gaiotto:2003yb} mathematically precise, but there are important differences that we will highlight.

\subsection{The Generalized Kontsevich Model Extended}
A topological closed string partition function $\tau(\Lambda)$ corresponding to topological gravity in two dimensions coupled to topological matter is captured by the generalized Kontsevich model \cite{Kontsevich:1992ti,WittenAlgebraic,Adler:1992tj,Itzykson:1992ya}:
\begin{equation}
\tau(\Lambda) = \frac{N(\Lambda)}{D(\Lambda)} \, ,
\end{equation}
where the numerator $N(\Lambda)$ and the denominator $D(\Lambda)$ read
\begin{eqnarray}
N(\Lambda) &=& \int [dM]_{N } e^{-\alpha \frac{1}{p+1} \Tr [ (M+ \Lambda)^{p+1}]_{\ge 2}} \label{numerator}
\nonumber \\
D(\Lambda) &=& \int [dM]_{N } e^{- \frac{\alpha}{2}  \sum_{k=0}^{p-1} \Tr M \Lambda^k M \Lambda^{p-1-k}} \label{denominator}
\end{eqnarray}
The square brackets with lower index $\ge 2$ indicate that one should consider only terms that are of order two or higher in the $N \times N$ hermitian matrix integration variable  $M$.
The coupling constant $\alpha$ can be identified with one over the string coupling, $\alpha=-1/g_s$. The partition function of the closed string is a function of the matrix source $\Lambda$ which codes the values of all couplings to the matter primaries and gravitational descendants. When the couplings go to zero, the matrix source $\Lambda$ goes to infinity, and the $\tau$ function approaches one.
The matter content is specified by the order $p+1$ of the potential term. We have that $p=2$ for pure gravity.\footnote{We remark that these matrix integrals should be thought off as consisting of a Gaussian integration plus an exponential of higher powers that are to be expanded as formal power series in the integration variable.
For a  careful treatment of these and other aspects see e.g. \cite{Kontsevich:1992ti, Buryak:2015eza}.}

Our set-up is elementary. We take the source matrix as well as the integration variable to be an $(N+1) \times (N+1)$ matrix and wish to interpret the last diagonal entry of the source matrix as (the dual of) an open string (or boundary insertion) modulus. 
The matrix integral only depends on the eigenvalues of the source matrix $\Lambda_z$, which we take to be diagonal: 
\begin{equation}
\Lambda_z = \mbox{diag} (\lambda_1,\dots,\lambda_N,z) \, . \label{sourceNplusOne}
\end{equation}
We used 
the special notation $z$ for the last diagonal entry, which we plan to single out. We have the corresponding closed string partition function:
\begin{eqnarray}
\tau(\Lambda_z) &=& \frac{N(\Lambda_z)}{D(\Lambda_z)} \, . \label{NplusOne}
\end{eqnarray}
To perform the integration over the $2N$ off-diagonal degrees of freedom, we integrate over  all $N (N+1)$ off-diagonal degrees of freedom, and then reinstate those that we wish to keep. To perform the integration, our main tool is  the Harish-Chandra-Itzykson-Zuber integration formula \cite{Harish-Chandra:1957dhy,Itzykson:1979fi}. We integrate in two steps. Firstly, we concentrate on the numerator, and then we perform the Gaussian integration in the denominator.

The exponent in the numerator  (\ref{numerator}) of equation (\ref{NplusOne})
can be simplified by shifting away the source $\Lambda_z$, to keep only a term which has power $p+1$, and a linear and constant term in the integration variable $M$. Indeed, we have
\begin{equation}
{[}(M+ \Lambda_z)^{p+1} {]}_{\ge 2} = (M+ \Lambda_z)^{p+1} -  \Lambda_z^{p+1} - (p+1)  \Lambda_z^p M \, ,
\end{equation}
and therefore upon shifting $M \longrightarrow M- \Lambda_z$, we find that the numerator $N(\Lambda_z)$ takes the  form:
\begin{align}
N(\Lambda_z) &=    \int [dM]_{N+1} e^{-\alpha \Tr \left( \frac{M^{p+1}}{p+1} - M \Lambda_z^{p} +\frac{p}{p+1}\Lambda_z^{p+1}\right)} \, .
\end{align}
We can then integrate over the unitary, angular variables that serve to diagonalise the matrix $M$. 
To that end we parameterize the $(N+1) \times (N+1)$ hermitian matrix $M$ in terms of a unitary matrix $U$ and diagonal matrix $M_d$ as $M= U M_d U^{-1}$. The matrix integral factorizes:
\begin{align}
N(\Lambda_z) &= \frac{\pi^{\frac{N(N+1)}{2}}}{\prod_{i=1}^{N+1} i!} \int \prod_{i=1}^{N+1}dm_i  \Delta_{N+1}(m) 
e^{-
\frac{\alpha}{p+1} \left( \sum_{i=1}^{N+1} m_i^{p+1} + p \Tr \Lambda_z^{p+1}\right)
}\cr
&\hspace{3cm}\times \int [dU]_{N+1} e^{\alpha \Tr \Lambda_z^p U M_d U^{-1}} \,,
\end{align}
where the integration variables $m_i$ are the eigenvalues of the diagonal matrix $M_d$ and we defined the Vandermonde determinant measure factor 
\begin{equation}
\Delta_N(m) = \prod_{1\le i < j \le N}(m_j-m_i)^2 \, .
\end{equation}
We  use the Harish-Chandra-Itzykson-Zuber formula \cite{Harish-Chandra:1957dhy,Itzykson:1979fi}
\begin{equation}
\int  [dU]_{N+1} e^{\alpha \Tr AUBU^{-1}} = \frac{\prod_{i=1}^N i!}{\alpha^{\frac{N(N+1)}{2}}} \frac{\text{det}(e^{\alpha\, a_i b_j}) }{\prod_{1\le i <j \le N+1} (a_j-a_i)(b_j-b_i)} \, ,
\end{equation}
to perform the integral over the unitary matrix $U$:
\begin{align}
N(\Lambda_z) &=\frac{1}{(N+1)!}\left(\frac{\pi}{\alpha}\right)^{\frac{N(N+1)}{2}}  \frac{1}{\prod_{1\le i < j \le N+1} (\lambda_{z,j}^p-\lambda_{z,i}^p)} e^{-\frac{\alpha p}{p+1} \Tr\Lambda_z^{p+1}}\cr
&\times \int \prod_{i=1}^{N+1} dm_i \, e^{-\frac{\alpha}{p+1} \sum_{i=1}^{N+1} m_i^{p+1}}\prod_{1\le i <j \le N+1} (m_j-m_i) \, \text{det}\left( e^{\alpha \lambda_{z,i}^p m_j } \right) \, .
\end{align}
In the next step, we isolate the integral over the last diagonal variable $m_{N+1}$ and write the remaining factors as an $N\times N$ matrix integral that we wish to identify as a generalized Kontsevich model with an extra insertion. The first step towards this goal is to expand the determinant  along the $(N+1)$st row of the matrix:
% XXX Check.
\begin{equation}
\det_{N+1} (e^{\alpha \lambda_{z,j}^p m_i}) \prod_{i<j}^{N+1} (m_j-m_i) = \sum_{l=1}^{N+1}  e^{\alpha z^p m_l} \det_N (e^{\alpha \lambda_{j}^p m_i})_{i \neq l}
\prod_{i<j \neq l}^{N+1} (m_j-m_i) \prod_{i \neq l}^{N+1} (m_l -m_i)  \, .
\end{equation}
By suitably changing variables in each term of the sum and using the permutation invariance of the measure, one can show that each resulting integral contributes equally.\footnote{We thank Alexandr Buryak for clarifying analogous steps that arise in \cite{Buryak:2015eza}.} Therefore, this sum may be written as $(N+1)$ times the contribution from the $l=N+1$ term. For easier writing, we rename the diagonal integration variable $m_{N+1}=s$ and find the numerator:
\begin{align}
N(\Lambda_z) &=\frac{1}{N!} \left(\frac{\pi}{\alpha}\right)^{\frac{N(N+1)}{2}} \frac{1}{\prod_{i=1}^N (z^p - \lambda_i^p)} e^{-\alpha\frac{p}{p+1} (z^{p+1} +\Tr\Lambda^{p+1})} \cr
& \int ds\, e^{-\alpha\frac{s^{p+1}}{p+1} + \alpha z^p\, s} \int \prod_{i=1}^N dm_i \Delta_N(m_i) \frac{\det(e^{\alpha \lambda_i^p m_j}) \prod_{i=1}^N (s-m_i)}{\prod_{1\le i < j \le N} (m_j-m_i) (\lambda_j^p - \lambda_i^p) } \, .
\end{align}
At this point, we trace our steps backwards and use the Harish-Chandra-Itzykson-Zuber formula in reverse.  Thus, we write the numerator in terms of the integration over a single variable $s$ and a $N\times N$ hermitian matrix $M$: 
\begin{align}
N(\Lambda_z) &= \left(\frac{\pi}{\alpha}\right)^N \frac{1}{\det(z^p - \Lambda^p)} \int ds\, e^{-\alpha\left(\frac{s^{p+1}}{p+1} - z^p\, s + \frac{p}{p+1} z^{p+1} \right) }\cr
&\times \int [dM]_N e^{-\alpha \Tr \left( \frac{M^{p+1}}{p+1} - \Lambda^p M + \frac{p}{p+1} \Lambda^{p+1} \right) } \, \det(s-M) \,.
\end{align}
This finishes the first and harder part of the calculation. Secondly,  we must keep track of the denominator $D(\Lambda_z)$ that serves to anchor the partition function $\tau$ at $1$ for large source matrix $\Lambda$.
The integration in the denominator is Gaussian, and the $N \times N$ determinant that will serve to normalize the final $N 
\times N$ matrix integration can easily be factored out:
\begin{align}
D(\Lambda_z) &= \int [dM]_{N+1} e^{- \frac{\alpha}{2}  \sum_{k=0}^{p-1} \Tr M \Lambda_z^k M \Lambda_z^{p-1-k}}
\nonumber \\
% &= \int\prod_{i,j} d m_{ij} e^{-\frac{\alpha}{2} \sum_{k=0}^{p-1} m_{ij} m_{ji} \frac{\lambda_{z,j}^p-\lambda_{z,i}^p}{\lambda_{z,j}-\lambda_{i,z}} }
% \nonumber \\
&= \prod_{i=1}^{N+1} \sqrt{\frac{2 \pi}{\alpha}} \frac{1}{\sqrt{p \lambda_{z,i}^{p-1}}}\prod_{i \neq j}^{N+1}\sqrt{ \frac{\pi}{\alpha}} \sqrt{\frac{\lambda_{z,j}-\lambda_{z,i}}{\lambda_{z,j}^p-\lambda_{i,z}^p} }
\nonumber \\
&=  \sqrt{\frac{2 \pi}{\alpha}}\, (p z^{p-1})^{-\frac{1}{2}} \prod_{i=1}^N \frac{\pi}{\alpha} \frac{\lambda_i -z}{\lambda_i^p-z^p}  \ D(\Lambda)
\nonumber \\
&= \sqrt{\frac{2 \pi}{\alpha}} \left(\frac{\pi}{\alpha}\right)^{N} \frac{\det (z-\Lambda)}{\det (z^p-\Lambda^p)} (p z^{p-1})^{-\frac{1}{2}}\  D(\Lambda) \, .
\end{align}
The integrating-out in both numerator and denominator provides us with a final formula for the tau-function at finite $N$:
\begin{align}
\tau(\Lambda_z) &= \sqrt{\frac{\alpha\, p}{2\pi}}\ \frac{z^{\frac{p-1}{2}}}{\det(z-\Lambda)} e^{-\alpha\frac{ p}{p+1} z^{p+1}} \int ds\, e^{-\frac{\alpha}{p+1}s^{p+1}+ \alpha \, z^p\, s} \cr
&\hspace{2cm} \times\frac{1}{D(\Lambda)} \int [dM]_N e^{-\alpha\frac{1}{(p+1)}\Tr \left [(M+\Lambda)^{p+1} \right]_{\ge 2} }\,   \text{det}(s-M-\Lambda) \,. 
\label{MainResult1}
\end{align}
This is our first technical result. To clarify its significance, we repackage it in various ways, and then provide an interpretation.

\subsection{The Closed and Open/Closed String Partition Functions}
To make contact with both the string theory and integrability literature, we slightly reshuffle 
the result \eqref{MainResult1}:
\begin{align}
\det\left(1-\frac{z}{\Lambda}\right)\,  \tau(\Lambda_z) 
&= \sqrt{ \frac{\alpha\, p}{2 \pi}}\,   z^{\frac{p-1}{2}}\,  e^{-\alpha\frac{p}{p+1} z^{p+1}}
\, \int ds\, e^{ \alpha\, z^p\, s}
\nonumber \\
&\hspace{1cm} \times  \frac{1}{D(\Lambda)} \int [dM] e^{-\alpha\frac{1}{(p+1)}\Tr \left [(M+\Lambda)^{p+1} \right]_{\ge 2} }\, e^{-\alpha\frac{1}{p+1} s^{p+1}}\,  \frac{\text{det}(\Lambda+M-s)}{\text{det}(\Lambda)} \, . \label{MainResult2}
\end{align}

\subsubsection*{The Closed String Partition Function}
We wish to be more specific about the interpretation and meaning of both the left and the right hand side of the equality (\ref{MainResult2}).
Firstly, let us remind the reader that the closed
string partition function $\tau(\Lambda)$ is the $\tau$-function of a reduced KP integrable hierarchy. The times $t_n$ of the integrable hierarchy are defined in terms of the source matrix $\Lambda$ by 
\begin{equation}
t_n = - \frac{1}{n}\, \Tr \Lambda^{-n} \, .
\label{definitiontimes}
\end{equation}
The times play the role of closed string couplings in topological  string theory. A well known and important result is the independence of the closed string partition function on the times $t_{np}$, where $n\in \mathbb{N}$ \cite{Kontsevich:1992ti, Itzykson:1992ya}. We will make use of this repeatedly in what follows. 

Secondly, we consider the closed string partition function $\tau(\Lambda_z)$ with the extra source eigenvalue $z$ (as in equation (\ref{sourceNplusOne})).
We can think of the addition of the extra source eigenvalue $z$
as a redefinition of the closed string couplings $t_n$. 
Indeed, we have:
\begin{equation}
\widetilde{t}_n = -\frac{1}{n}\,  \Tr \Lambda_z^{-n}
=-\frac{1}{n}\, \Tr \Lambda^{-n}  - \frac{1}{n\, z^{n}}
= t_n - \frac{1}{n\, z^{n}} \, .
\end{equation}
Thus, if we view the $\tau$-function as a function of the time variables $t_n$, which are independent in the large $N$ limit, then we can define a shifted closed string partition function by the formula:
\begin{equation}
\tau(\Lambda_z) %= G_z (\tau(t_{n})) 
= \tau\left(t_{n} - \frac{1}{n\, z^{n}}\right) \, .
\end{equation}
The extra source term redefines the closed string background. This elementary aspect of the  duality formula (\ref{MainResult2}) is well-understood in the  literature.

Let us  introduce an alternative symbol for the determinant on the left hand side of equation (\ref{MainResult2}):  
\begin{equation}
e^{\xi} = \det\left(1-\frac{z}{\Lambda}\right) \, ,
\end{equation}
which implies that 
\begin{align}
\label{xidefn}
\xi &= \sum_{n=1 %, n \notin p \mathbb{Z}
}^{\infty} z^n t_n \, .
\end{align}
We then recognize on the left hand side of equation (\ref{MainResult2}) a quantity from integrable systems \cite{Bertola:2014yka}, namely the extended partition function or wave potential $\tau_{\ext}(\Lambda, z)$ equal to:
\begin{equation}
\tau_{\ext}(\Lambda, z) 
= e^{\xi}\, \tau(\Lambda_z) \, .
\label{wavepotential}
\end{equation}
It is important that the extended tau-function $\tau_{\ext}(\Lambda_z)$ (in contrast to the tau-function $\tau(\Lambda)$) depends on all times of the KP integrable hierarchy, and in particular on the times $t_{np}$. The factor  $e^{\xi}$  introduces an exponential  $t_{np}$ dependence.
Thus the left-hand side $\tau_{\ext}(\Lambda_z)$ of the equality (\ref{MainResult2}) is an extended and shifted closed string partition function, equal to the wave potential of the KP integrable hierarchy.

\subsubsection*{The Wave Potential and the Open String Partition Function}
\label{MainTheorem}

Moreover, we also wish to compactly code and interpret the right hand side of equation \eqref{MainResult2}. To that end we define an open/closed partition function:
\begin{align}
\tau^{\op+\cl} (\Lambda,s) 
&= \frac{1}{D(\Lambda)} \int [dM] e^{-\alpha\frac{1}{(p+1)}\Tr \left [(M+\Lambda)^{p+1} \right]_{\ge 2} }\, e^{-\alpha\frac{1}{p+1} s^{p+1}}\,  \frac{\text{det}(\Lambda+M-s)}{\text{det}(\Lambda)} \, ,
\label{definitiontauopcl}
\end{align}
which is equal to the generalized Kontsevich matrix integral with a  normalized determinant insertion labelled by a coupling constant $s$. Our result (\ref{MainResult2}) then reads more compactly:
\begin{align}
\tau_{\ext}(\Lambda, z) &= 
\sqrt{ \frac{\alpha\, p}{2 \pi}} %(\det(z^p-\Lambda^p))^{-1/p} 
 z^{\frac{p-1}{2}}\,   e^{ -\alpha \frac{p}{p+1} z^{p+1}} 
 \int ds\, e^{  \alpha\, z^p\, s}\, \tau^{\op+\cl}(\Lambda,s)
 \, .
\end{align}
If we moreover define a formal Fourier transform of a function (or rather a formal power series) $f(s)$ by:
\begin{align}
\Phi[f(s)](z)& := \sqrt{\frac{\alpha\, p}{2 \pi}}\,  z^{\frac{p-1}{2}} \int ds\, e^{\alpha \left( \frac{z^{p+1}}{p+1} + z^p s\right)} \, f(s+z) \,,
\end{align}
we  have that our duality succinctly reads:
\begin{align}
\tau_{\ext}(\Lambda, z) &= \Phi\left[ \tau^{\op+\cl}(\Lambda,s) \label{MainResult3}
\right](z) \,.
\end{align}
We summarize that the (finite $N$ generalization of the) extended closed string partition function equals the integral transform of the open/closed string partition function. 
Importantly, we still need to  argue that the correct interpretation of the right hand side of the result (\ref{MainResult3}) is indeed as an open/closed string partition function. In order to do so, we need considerably more background.

Relatively recently, the following wave function was introduced in the integrable system literature \cite{Buryak:2014dta,Bertola:2014yka}.
Given a $p$-reduced integrable KP hierarchy and the associated Lax operator $L$  -- see \cite{Dickey:1991xa}
for a pedagogical introduction to the subject --
satisfying the evolution equations: 
\begin{equation}
\frac{\partial L}{\partial t_m} = [ [L^{\frac{m}{p}}]_{\ge 0}\,, L ] \, ,
\end{equation}
one defines a wave function $\Psi(t_k)$ that satisfies the differential equations:
\begin{equation}
\frac{\partial \Psi}{\p t_m} = [L^{\frac{m}{p}}]_{\ge 0}\, \Psi \, ,
\label{WaveEquations}
\end{equation}
as well as the initial condition  \cite{Buryak:2014dta,Bertola:2014yka}
\begin{equation}
\Psi_{t_{n \ge 2}=0}=1 \,. \label{InitialCondition}
\end{equation}
The interest in the wave function $\Psi(t_k)$ originates in the advance that has been made in computing intersection numbers on moduli spaces of Riemann surfaces with boundaries
\cite{Buryak:2014dta,Buryak:2018ypm}. Indeed, in the presence of boundaries, the reduction of the integrable hierarchy no longer takes place, and the generating function of intersection numbers does depend on the times $t_{np}$. Moreover, there is an extra dependence on a coupling constant $s$ that counts open string or boundary insertions.
The partition function  of closed/open intersection numbers can  be expressed in terms of the wave function $\Psi$ \cite{BCT1,Buryak:2018ypm}. Let us describe the final result of the algebro-geometric calculations. We introduce  free energies that generate intersection numbers on Riemann surfaces with and without boundary: 
\begin{align}
\tau &=  e^{F^{\cl}}
\nonumber \\
\tau^{\op+\cl}_{\text{geom}} &=  e^{ F^{\cl} +F^{\op}} =   \tau \, e^{F^{\op}} 
\, .
\end{align}
% g=2h+b-1 
Crucially, it is shown in \cite{Buryak:2018ypm} that the generator $F^{\text{open}}$ of geometric open/closed intersection numbers is given by the logarithm of the wave function $\Psi(t_k)$ after a substitution of variables:
\begin{equation}
F^{\op} = 
\log \Psi  
\left({t}_i,t_p \rightarrow 
t_p-\alpha s \right)
\, .
\end{equation}
It is important to note here that in the generating function, we have allowed for extended closed string amplitudes, namely, amplitudes where we have added boundaries but no explicit open string insertions. In this regard, see the useful intuitive remarks in  \cite{BCT1}.

Finally, we are ready to state the relation our matrix model has to the algebro-geometric open/closed intersection numbers.  We propose that our matrix integral $\tau^{\op+\cl}$ is equal to its geometric counterpart:
\begin{eqnarray}
\tau^{\op+\cl} = \tau^{\op+\cl}_{\text{geom}} = \tau(t) \, \Psi({t},t_p- \alpha s ) 
\, .\label{GeometricMainResult}
\end{eqnarray}
The left hand side is our open/closed string partition function  $\tau^{\op+\cl}$ defined in terms of the generalized Kontsevich integral with normalized determinant insertion \eqref{definitiontauopcl}, while the right hand side is the generator of geometric open/closed string intersection numbers, according to  \cite{Buryak:2018ypm}. 

We will prove the identification \eqref{GeometricMainResult} in the following. As a warm-up, 
let us first show that our $\tau^{\op+\cl}$  indeed only depends on the combination $t_p-\alpha s$ as expected from equation (\ref{GeometricMainResult}).
We write the open/closed partition function $\tau^{\op+\cl}(\Lambda,s)$ as the inverse integral transform of the wave potential $\tau_{\ext}(\Lambda,z)$:\footnote{To obtain this expression, the contour in the $z$-plane is engineered to give the equality 
\begin{equation}
\frac{\alpha p}{2\pi} \int dz\, z^{p-1}\, e^{\alpha z^{p} (s-s')} =\frac{1}{2\pi} \int d(\alpha z^{p})\, e^{\alpha z^{p} (s-s')} = \delta(s-s')
\end{equation}}
\begin{align}
\tau^{\op+\cl}(\Lambda, s) =  \sqrt{\frac{\alpha\, p}{2 \pi}} \int\, dz\, z^{\frac{p-1}{2}} e^{\alpha \frac{p}{p+1} z^{p+1} -\alpha\, z^p\, s} \tau_{\ext}(\Lambda,z) \, . \label{tauopcltauext}
\end{align}
From this expression, one verifies that the open/closed partition function is annihilated by the following operator:
\begin{align}
\left(\frac{1}{\alpha}\frac{\p}{\p s} + \frac{\p}{\p t_p} \right) \tau^{\op+\cl}(\Lambda, s) &= 
\sqrt{\frac{\alpha\, p}{2 \pi}} \int\, dz\, z^{\frac{p-1}{2}} e^{\alpha \frac{p}{p+1} z^{p+1} -\alpha\, z^p\, s} \left(- z^p +\frac{\p}{\p t_p} \right)\tau_{\ext}(\Lambda,z) \,.\cr
&= 0 \,.
\end{align}
In the second equality, we have used that the dependence of the wave potential on the time $t_p$ comes purely from the $e^{\xi}$ factor as the purely closed string partition function $\tau(\Lambda_z)$ is independent of the times $t_{np}$. This finishes our proof of the elementary property.

More importantly, we will prove that the wave function $\Psi(t_k)$ which is equal to the ratio of our matrix model integral $\tau^{\op+\cl}$ and the original closed string tau-function $\tau(\Lambda)$ does indeed satisfy the differential equations \eqref{WaveEquations} and the initial conditions \eqref{InitialCondition} in Section \ref{Proof}. Since the solution to these equations is unique, that will prove our claim \eqref{GeometricMainResult}.

In closing we remark that the matrix model perspective on both the algebraic and stringy duality naturally gives rise to the extended closed string partition function encountered in \cite{BCT1}. This is directly related to the wave function that is canonical from the perspective of  the integrable system \cite{Buryak:2014dta,Bertola:2014yka}, i.e. it is the wave function that governs the algebraic intersection numbers \cite{Buryak:2018ypm}, including their extra time dependencies.

\section{The Open/Closed String Virasoro Constraints}
\label{Virasoro}
The KdV integrable hierarchy combined with one Virasoro constraint -- the string equation -- is sufficient to determine all correlators of topological gravity coupled to matter. Equivalently, the W-algebra constraints on topological gravity correlators are sufficient to determine them all. A subset of the W-algebra constraints are the Virasoro equations, which are strong constraints on the partition function and which allow to make clear contact with the free fermion formulation of the integrable hierarchy, for instance.  They will also allow us to make further contact with the geometric framework \cite{Buryak:2018ypm}. Crucially, they are instrumental  in proving the identification between our matrix integral and the generator of geometric invariants. The identification of these Virasoro algebras goes back to the study of open/closed string matrix models  \cite{Dalley:1992br,Johnson:1993vk} and has also been touched upon in \cite{Gaiotto:2003yb,Dijkgraaf:2018vnm}, amongst many other places. We hope our treatment clarifies the various guises of the Virasoro algebra in the literature. 

\subsection{The Open/Closed Virasoro Algebra}
\label{extendedVirasoro}

The closed string partition function is annihilated by half of a Virasoro algebra  \cite{Fukuma:1990jw,Dijkgraaf:1990rs,Adler:1992tj,Kharchev:1991cy,Dijkgraaf:1991qh}:
\begin{equation}
L_{n} \tau(\Lambda_z) = 0 \qquad\text{for}\qquad n\ge -1\,.
\end{equation}
The generators of the Virasoro algebra are given traditionally in terms of the times $t_k$ of the integrable system:
\begin{align}
\label{oldVirasoro}
L_{-1} &= \alpha \frac{\p}{\p t_1} + \sum_{k=p+1}^{\infty\ \prime} \frac{k}{p} t_k \frac{\p}{\p t_{k-p}} +\frac{1}{2p}\sum_{k=1}^{p-1}k(p-k)\, t_k t_{p-k} \cr
L_0 &= \alpha \frac{\p}{\p t_{p+1}}+   \sum_{k=1}^{\infty\ \prime} \frac{k}{p} t_k \frac{\p}{\p t_{k}} + \frac{p^2-1}{24p} \cr
L_{n} &= \alpha \frac{\p}{\p t_{(n+1)p+1}} + \sum_{k=1}^{\infty\ \prime}\frac{k}{p}t_k \frac{\p}{\p t_{np+k}} + \frac{1}{2p}  \sum_{k=1}^{pn-1\ \prime} \frac{\p^2}{\p t_{k} \p t_{np-k} }  \,. \cr
\end{align}
The primed sums have indices in the reduced set of times $t_{k \notin p \mathbb{N}}$.
Here we study a set of extended Virasoro generators $L_n^{\ext}$ that act on the wave potential $\tau_{\ext}(\Lambda,z)$ in such a manner that they become differential operators in the spectral parameter $z$ only. Moreover, after Fourier transform, they become differential operators in the variable $s$. Thus, we will find  the following equalities: 
\begin{equation}
\label{openclosedLL}
L_n^{\ext}(t_k) \tau_{\ext} = -L_n^{\ext}(z)\Phi[ \tau^{\op+\cl}(s)](z) = - \Phi[ L^{\ext}_n(s) \tau^{\op+\cl} ] \,. 
\end{equation}
Therefore, the wave potential $\tau_{\ext}$ will be annihilated by the generators
\begin{equation}
L_n^{\cl} = L_n^{\ext}(t_k) + L_n^{\ext}(z) \label{ClosedVirasoro} \, ,
\end{equation}
while the open/closed partition function $\tau^{\op+\cl}$ will be killed by the Virasoro generators:
\begin{equation}
L_n^{\op} = L_n^{\ext}(t_k) + L_n^{\ext}(s) \, . \label{OpenVirasoro}
\end{equation}
 To prove the equalities \eqref{openclosedLL}, we firstly propose that the extended Virasoro generators are given by the following expressions in terms of the times $t_k$: 
\begin{align}
L_{-1}^{\ext} &= \alpha \frac{\p}{\p t_1} + \sum_{k=p+1}^{\infty} \frac{k}{p} t_k \frac{\p}{\p t_{k-p}} +\frac{1}{2p}\sum_{k=1}^{p-1}k(p-k)\, t_k t_{p-k} +t_p\cr
L_0^{\ext} &= \alpha \frac{\p}{\p t_{p+1}}+   \sum_{k=1}^{\infty} \frac{k}{p} t_k \frac{\p}{\p t_{k}} + \left(\frac{p^2-1}{24p} +\frac{1}{2p}\right)\cr
L_{n}^{\ext} &= \alpha \frac{\p}{\p t_{(n+1)p+1}} + \sum_{k=1}^{\infty}\frac{k}{p}t_k \frac{\p}{\p t_{np+k}} + \frac{1}{2p}  \sum_{k=1}^{pn-1} \frac{\p^2}{\p t_{k} \p t_{np-k} } + \, \frac{1}{p} \frac{\p}{\p t_{np}} \,. \cr
\label{extendedClosedVirasoro}
\end{align}
Note that all times $t_{np}$ appear in these extended Virasoro generators. 
This is a natural change, since the wave potential depends on all times. Secondly, we observe further additional terms that can formally be argued as follows. If we were to introduce a time $t_0$ such that $t_0 = -\frac{1}{0} \text{Tr} \Lambda^0$ -- compare to equation \eqref{definitiontimes} --, then $0 \, t_0$ would measure minus the dimension of the matrix $\Lambda$. Since the latter changes by one in our process of integrating out off-diagonal degrees of freedom, the combination $0 \, t_0$  can be argued to change by one. This formally gives rise to the last term in the first and third lines of the proposal (\ref{extendedClosedVirasoro}).
The additional contribution to the constant term in $L_0$ can then be obtained by computing the $[L^{\ext}_1, L^{\ext}_{-1}]$ commutator. A more solid argument for the proposal is the long calculation that follows.

By explicitly acting with the extended Virasoro generators $L_n^{\ext}(t_k)$ on the wave potential $\tau_{\ext}$, we will manage to represent the extended Virasoro operators as differential operators in $z$ exclusively. 
Firstly, we calculate  the action of the  generators $L_{n>1}^{\ext}$ on the wave potential:
\begin{align}
L_n^{\ext}  \tau_{\ext} &= L_n^{\ext}  \left(e^{\xi}\, \tau(\Lambda_z) \right) \cr
& = \tau(\Lambda_z)\, L_n^{\ext}  e^{\xi}  +  e^{\xi}\, L_n^{\ext}  \tau(\Lambda_z) + 
\frac{1}{p} \sum_{k=1}^{pn-1} \frac{\p e^{\xi}}{\p t_{np-k}} \frac{\p \tau(\Lambda_z)}{\p t_{k} }\,.
\label{Lnaction}
\end{align}
We have taken into account the cross term that arises due to the double-derivative term in the $L^{\ext}_{n>0}$ generators. Let us consider each of these terms in turn.
The first term is proportional to:
\begin{equation}
L_n^{\ext}  e^{\xi} = \left( \alpha\, z^{np+p+1} + \frac{1}{p}\sum_{k=1}^{\infty}k t_k z^{np+k} + \frac{1}{2p} \sum_{k=1}^{np-1} z^{np} + \frac{1}{p}z^{np} \right)e^{\xi} \, . \label{intermediate}
\end{equation}
Observe that the second term in equation (\ref{intermediate}) can be rewritten as a $z$-derivative:
\begin{equation}
\label{Lnfinal}
L_n^{\ext}\,  e^{\xi} = \left( \alpha\, z^{np+p+1} + \frac{z^{np+1}}{p}\frac{\p}{\p z}+ \frac{np+1}{2p} z^{np}  \right)e^{\xi} \, .
\end{equation}
Let us now consider the second term in equation \eqref{Lnaction}. The important input here is that the closed string partition function $\tau(\Lambda_z)$ is annihilated by the action of the shifted Virasoro algebra where the shifted times $\widetilde{t}_k$ are given by
\begin{equation}
\widetilde{t}_k  = t_k - \frac{1}{k\, z^k} \,.
\label{shiftedtimes}
\end{equation}
The shifted Virasoro generators take the simple form:
\begin{equation}
L_n^{\ext}(t_k) = L_n^{\ext}(\widetilde{t}_k) +  \frac{1}{p}\sum_{k=1}^{\infty} \frac{1}{z^k} \frac{\p }{\p t_{np+k}}\,.
\end{equation}
Acting on $\tau(\Lambda_z)$ the first term gives a zero contribution while the second term can be rewritten as:
\begin{align}
L_n^{\ext}\,  \tau(\Lambda_z) &= 
 \frac{1}{p}\, z^{np+1}\, \frac{\p \tau^{\cl}(\Lambda_z)}{\p z} - 
 \frac{1}{p}\, z^{np+1}\,  \sum_{k=1}^{np-1}z^{-k-1}\frac{\p \tau(\Lambda_z)}{\p t_{k}}\,.
\label{Lnontau}
\end{align}
Finally, the  last term in equation \eqref{Lnaction} can be evaluated to be: 
\begin{equation}
\frac{e^{\xi}}{p} \, z^{np+1} \sum_{k=1}^{np-1} z^{-k-1} \frac{\p \tau(\Lambda_z)}{\p t_{k} }\,.
\end{equation}
We see that this contribution  cancels the second term in equation \eqref{Lnontau} after multiplication by $e^{\xi}$. Furthermore, the coefficients of the $\p / \p z$ terms in
formulas \eqref{Lnfinal} and \eqref{Lnontau} are the same. Therefore, summing over all terms, we find the following  differential operator as the action of the extended Virasoro generators on the wave potential:
\begin{align}
\label{Lclz}
L_{n}^{\ext}(t_k)  \tau_{\ext}   
%&= (-1)^n \frac{(\alpha\, z^p)^n}{p} \left[ z\frac{\p }{\p z} +\Sigma+ \alpha\, p\,  z^{p+1} \right] \cr
& = \left(\alpha z^{np+p+1}   + \frac{1}{p}  z^{np+1} \frac{\p}{\p z} + \frac{ np+1}{2p} \, z^{np} \right)  \tau_{\ext} \,.
\end{align}
The differential operator on the right hand side is the negative of what we defined to be $L_n^{\ext}(z)$ in equation \eqref{ClosedVirasoro} for the $n>0$ Virasoro generators. We turn  to the remaining two cases. The calculations for $L_0$ are  similar to the ones we have already done and we find
\begin{equation}
L_{0}^{\ext}  \tau_{\ext} = \left(\alpha\, z^{p+1}+\frac{z}{p}\frac{\p}{\p z} + \frac{1}{2p} \right)\tau_{\ext}
% = \alpha\, z^p\left(z + \frac{1}{\alpha\, p\, z^{p -1}} \frac{\p}{\p z}  - \frac{p-1}{2\alpha\, p\, z^p}\right)\tau_{\ext} + \frac{1}{2}\, \tau_{\ext}
\,.
\end{equation}
Note that the constant $1/(2p)$ term survives in the first equality because the shifted Virasoro generator $L_0(\widetilde{t}_k)$ that annihilates the closed string partition function $\tau(\Lambda_z)$ does not include this constant term. 

The analysis for the operator $L_{-1}^{\ext}$ is more involved due to the  terms quadratic in the times. We use again the fact that the operator $L_{-1}^{\ext}(\widetilde{t}_k)$, with shifted times, annihilates the closed string partition function and find the following relation:
\begin{align}
L_{-1}^{\ext}   \tau_{\ext}&= \left(\alpha\, z+\sum_{k=p+1}^{\infty}\frac{k}{p}t_kz^{k-p} \right) \tau_{\ext} + e^{\xi}\, \sum_{k=p+1}^{\infty} \frac{1}{p} \frac{1}{z^k} \frac{\p \tau}{\p t_{k-p}}\cr
&~+  \left( t_p  + \frac{1}{2p} \sum_{k=1}^{p-1} k(p-k)\left( \frac{t_{p-k}}{k z^k} + \frac{t_k}{(p-k) z^{p-k}} - \frac{1}{k(p-k)z^p} \right)   \right) \tau_{\ext}\,.\cr
\end{align}
We  combine  terms linear in the times $t_k$ and   use the equations
\begin{equation}
\sum_{k=1}^{\infty} \frac{k}{p}\, t_k\, z^{k-1} e^{\xi} = \frac{1}{p} \frac{\p e^{\xi}} {\p z} \qquad \mbox{and} \qquad
\sum_{k=1}^{\infty} z^{-k-1} \frac{\p \tau (\tilde{t}_k)}{\p t_k} = \frac{\p \tau(\tilde{t}_k)}{\p z}\, ,
\end{equation}
to finally obtain
\begin{align}
\label{Lminusoneontauext}
L_{-1}^{\ext}\,   \tau_{\ext}
%&= \left( -z - (-\alpha)^{-1}\frac{(p-1)}{2p z^p} +(-\alpha)^{-1} \frac{1}{p z^{p-1}} \frac{\p}{\p z}  \right) \tau_{\ext}\cr
&= \alpha\left(z + \frac{1}{\alpha\, p\, z^{p-1}} \frac{\p}{\p z}  -\frac{p-1}{2\alpha\, p\, z^p } \right)  \tau_{\ext}\,.
\end{align}
We note that the $t_p$ term in the extended Virasoro generator
\eqref{extendedClosedVirasoro} is crucial in order to obtain the sum over all times that leads to the $\p_z$-derivative acting on the $e^{\xi}$ factor. In short, the form
of the Virasoro generators \eqref{Lclz} is valid for all $n$.
This concludes our analysis of the extended closed Virasoro generators and how they are represented as differential operators in $z$ on the wave potential.  We surmise that indeed the closed string Virasoro generator $L_n^{\cl}$ in equation \eqref{ClosedVirasoro} annihilates the wave potential $\tau_{\ext}$.

\subsection{From Closed to Open Virasoro}

\label{ClosedOpenVirasoro}
Our next step in proving the equalities \eqref{openclosedLL} is to Fourier transform the $z$-differential operator using the open/closed duality equation \eqref{MainResult3} to find the  open string realization of the Virasoro algebra. To make things more transparent, we first rewrite the extended Virasoro algebra in \eqref{Lclz} in the following  manner:
\begin{align}
\label{Lclz2}
L_{n}^{\ext}  \tau_{\ext} 
%&=  \alpha\, z^{p(n+1)}\left(z + \frac{1}{\alpha\, p\, z^{p-1}} \frac{\p}{\p z}  \right) + \frac{np+1}{2p}  z^{np} \cr
&= \alpha\, z^{p(n+1)}\left(z + \frac{1}{\alpha\, p\, z^{p-1}} \frac{\p}{\p z}  -\frac{p-1}{2\alpha\, p\, z^p } \right) + \frac{n+1}{2}  z^{pn} \,.
\end{align}
We now make use of the properties of the Fourier transform that are proven in  Appendix \ref{FourierTransform} (see equation \eqref{GFTprop12}) that effectively show the equivalence between differential operators of $z$ acting on the closed string side and differential operators of $s$ acting on the open string side. We find the  Virasoro generators: 
\begin{align}
% L_{-1}^{\ext}(s) &= -\alpha\, s \cr
% L_{0}^{\ext}(s) &=  \frac{\p}{\p s} \cdot  s - \frac{1}{2}= s\frac{\p}{\p s}+\frac{1}{2}\cr
L_n^{\ext}(s) &=  (-\alpha)^{-n} \left( \frac{\p^{n+1}}{\p s^{n+1}} \cdot s -  \frac{1}{2}(n+1) \frac{\p^n}{\p s^n} \right)\cr
& =  (- \alpha)^{-n} \left( s\frac{\p^{n+1}}{\p s^{n+1}} +  \frac{1}{2}(n+1) \frac{\p^n}{\p s^n} \right) \, . \label{sVirasoro}
\end{align}
Thus, we have shown that the operators $L_n^{\op}$ defined in equation \eqref{OpenVirasoro} as the sum of the extended Virasoro generators and the $s$ differential operators  annihilate the open/closed partition function  $\tau^{\op+\cl}$.

Let us briefly remark on the string coupling dependence of the Virasoro generators. 
The terms in equation \eqref{extendedClosedVirasoro} become more transparent when we rescale the times by a factor of $\alpha=-1/g_s$. The resulting string coupling dependence is then $g_s^{-2}$ for the term quadratic in times in $L_{-1}^{\ext}$, arising from the sphere,
and there is an extra term arising from the disk, proportional to $g_s^{-1}$. 
The term with the two derivatives in $L_n^{\ext}$ is proportional to $g_s^2$, and the last, single derivative term is linear in the string coupling $g_s$. The terms above, in equation \eqref{sVirasoro} are proportional to $g_s^n$. Each open string insertion comes with an extra factor of the string coupling $g_s$.

\subsubsection*{An Alternative Point of View}
Let us add a remark on how the open/closed Virasoro generators are related to
those found in the literature for $p=2$. See e.g. \cite{Buryak:2015eza,Dijkgraaf:2018vnm}.  We show in Appendix \ref{AlternativeVirasoro} that those Virasoro  generators 
(which we also compute in the Appendix  for general $p$) are related to the ones we have by conjugation.
Indeed, we can conjugate the Virasoro generators to eliminate
the $t_{np}$ dependence of the wave potential $\tau_{\ext}$ on which they act. 
The open/closed Virasoro generators then match the  generators 
of references 
\cite{Buryak:2015eza} and \cite{Dijkgraaf:2018vnm} that study the case $p=2$.
It should be remarked however (see appendix \ref{AlternativeVirasoro}) that this point of view introduces either an intricate operator insertion  $\det ( \partial_s/\alpha + \Lambda^p)^{-\frac{1}{p}}$ inside the matrix expression for the open string partition function, or  the ratio of determinants insertion discussed in appendix \ref{PureAlternative} for $p=2$  \cite{Buryak:2015eza}. Since our matrix model perspective in the bulk of the paper naturally matches both the algebro-geometric considerations as well as the integrability literature, we have stuck to this point of view in section \ref{matrixintegral} as well as in the calculation of the open/closed Virasoro generators presented in subsections \ref{extendedVirasoro} and \ref{ClosedOpenVirasoro}. Finally, we remind the reader that these Virasoro algebras have an interpretation in terms of a primary operator insertion that dates back to \cite{Dalley:1992br,Johnson:1993vk}.

\subsection{The Integrable Hierarchy and the Relation to Geometry}
\label{Proof}
In this subsection, we tie up several loose ends. We first make good use of the operators we found in our analysis to prove the statement made in subsection 
\ref{MainTheorem} that our open/closed partition function matches the algebro-geometric generating function. Secondly, we note that a subset of our Virasoro constraints correspond to  geometric constraints derived in \cite{Buryak:2018ypm}, and that our Virasoro algebra extends those constraints to an infinite family.

\subsubsection*{The Connection to Algebraic Geometry}
The first part of the proof of the fact that our matrix partition function matches the geometric one is based on a classic result in integrable systems, which says that the Baker-Akhiezer wave function $\psi(t_{k},z)$ is given by:
\begin{equation}
\psi(t_{k},z) = \frac{\tau(\widetilde{t}_{k})}{\tau(t_k)} e^{\xi(t_k,z)}
= \frac{\tau_{\ext}}{\tau} \, , \label{BA}
\end{equation}
where $\widetilde{t}_k$ are the shifted times \eqref{shiftedtimes}.
The Baker-Akhiezer wave function $\psi(t_{k},z)$ satisfies the differential equations \eqref{WaveEquations} as well as  the eigenvalue equation
\begin{equation}
L\, \psi(t_{k},z) = z^p\, \psi(t_{k},z) \, , 
\label{BAeigenvalue}
\end{equation}
where $L$ is the Lax operator of the integrable hierarchy. Moreover, from the linear relation between the open/closed partition function $\tau^{\op+\cl}$ and the wave potential $\tau_{\ext}$ in \eqref{tauopcltauext}, it is clear that the ratio
$\tau^{\op+\cl}/\tau$ is related to the Baker-Akhiezer function $\psi(t_k,z)$ via the Fourier transform. Therefore the ratio
$\tau^{\op+\cl}/\tau$ also satisfies the differential equations \eqref{WaveEquations}. 

It  remains to prove that the wave function $\Psi(t_k) =\tau^{\op+\cl}/\tau$ also satisfies the initial conditions \eqref{InitialCondition}. Our proof closely follows the  proof in \cite{Bertola:2014yka}. Previously, we proved that the wave potential satisfies the equation (see \eqref{Lminusoneontauext}):
\begin{equation}
\label{LminusoneSz}
L_{-1}^{\ext}\cdot \tau_{\ext} = \alpha\, S_z\cdot \tau_{\ext}\,,
\end{equation}
where we define $S_z$ to be the differential operator:
\begin{equation}
S_z = \left(z + \frac{1}{\alpha\, p\, z^{p-1}} \frac{\p}{\p z}  -\frac{p-1}{2\alpha\, p\, z^p } \right) \,.
\end{equation}
%
% Now, the wave potential is related to the Baker-Akhiezer wave function by the relation
% %
% \begin{equation}
% \psi(t_k, z) = \tau(t_k)\, \tau_{\ext}(t_k, z) \,.
% \end{equation}
%
We now set all times to zero except the time $t_1$ and study the reduced Baker-Akhiezer function $\left.\psi(t_k, z) \right|_{t_{\ge 2}=0}$. When all times except time $t_1$ are zero, the closed string partition function $\tau(t)$ equals one. Also, recall that the operator $L_{-1}^{\ext}$ coincides with the operator $\alpha\, \partial_{t_1}$ when all times but the first are zero.\footnote{These two statements have to be modified for $p=2$. For this value of $p$, the partition function at zero higher times equals $\tau(t)=\exp \frac{t_1^3}{6}$, and the operator $L_{-1}^{\ext}$ has an extra term $t_1^2/2$. These two modifications cancel each other in the reasoning.} Combining this fact with equation \eqref{LminusoneSz} and the identification \eqref{BA}, we conclude that \cite{Bertola:2014yka}
\begin{equation}
\left(\frac{\p }{\partial_{t_1}}\right)^k \left.\psi(t_k, z) \right|_{t_{\ge 2}=0} =  S_z^k \, \left.\psi(t_k, z) \right|_{t_{\ge 2}=0}
\end{equation}
If we therefore define the function $A(z)$ to be the value that the wave function takes at $t_1=0$, 
\begin{equation}
\left.\psi(t_k, z) \right|_{t_{\ge 1}=0} = A(z)\,,
\end{equation}
then we see that by  Taylor expansion, we have
\begin{align}
\label{wavefunctiondefn}
\left. \psi(t_k, z)\right|_{t_{\ge 2}=0} =\sum_{n=0}^\infty \frac{1}{n!}\, S_z^n\, A(z)\, t_1^n \,.
\end{align}
Moreover, one can check that the function $A(z)$ satisfies the differential constraint  \cite{Bertola:2014yka}:
\begin{equation}
\label{Azdefn}
S_z^p\cdot A(z) = z^p\, A(z)\,. 
\end{equation}
This follows  from firstly, the property \eqref{BAeigenvalue}
of the Baker-Akhiezer function that it is an eigenvector of the Lax operator, secondly, the initial  conditions that require the Lax operator $L$ to be $(\partial _{t_1})^p$ when all times are zero and thirdly, the $L_{-1}^{\ext}$ Virasoro identity \eqref{LminusoneSz}. 
We now claim that  the function $A(z)$ that satisfies  the differential constraint
\eqref{Azdefn} is given by the following contour integral:
\begin{equation}
\label{Azsolution}
A(z) =   \sqrt{\frac{\alpha\, p}{2 \pi}}\, z^{\frac{p-1}{2}}\, e^{-\alpha \frac{p}{p+1} z^{p+1}}\, \int ds e^{-\alpha \frac{s^{p+1}}{p+1} + \alpha\, s\, z^p}
\, .
\end{equation}
We check by explicitly acting with the operator $S_z$:
\begin{equation}
S_z^p\cdot A(z) = \sqrt{\frac{\alpha\, p}{2 \pi}}\, z^{\frac{p-1}{2}}\, e^{-\alpha \frac{p}{p+1} z^{p+1}}\, \int ds \, s^p\, e^{-\alpha \frac{s^{p+1}}{p+1} + \alpha\, s\, z^p}\,.
\end{equation}
Within the $s$-integral, one adds and subtracts $z^p$. Then, the combination $(s^p - z^p)$ can be written as a total derivative, which integrates to zero. We have thus proven \eqref{Azdefn}.  We have chosen the normalization of the function 
\eqref{Azsolution} such that it is the inverse Fourier transform of the constant function equal to $1$. 

Recall  that our wave function $\Psi(t_k) = \tau^{\op+\cl}/\tau$ is the formal Fourier transform of the Baker-Akhiezer function  \eqref{wavefunctiondefn}:
\begin{equation}
\left. \Psi(t_k)\right|_{t_{\ge 2}=0} =\sqrt{\frac{\alpha\, p}{2 \pi}}\int dz\, z^{\frac{p-1}{2}}\, e^{\alpha\, p \frac{z^{p+1}}{p+1} - \alpha\, z^p\, t_p}\, \psi(t_k, z)_{t_{\ge 2}=0} \label{FTBA}
\end{equation}
The key point is now that every term in the sum \eqref{wavefunctiondefn} for $n>0$ vanishes  inside the integral \eqref{FTBA} \cite{Bertola:2014yka}. This can be  checked by integration by parts. 
%
%\begin{equation}
%\left(-\frac{1}{\alpha\, p}\frac{\p}{\p %z}\cdot z^{1-p}  -\frac{p-1}{2\,\alpha\, p\, %z^p} + z \right) z^{\frac{p-1}{2}}\, %e^{\alpha\, p \frac{z^{p+1}}{p+1} - \alpha\, %z^p\, t_p}= 0\,.
%\end{equation}
%
Therefore the only term that contributes to $\left. \Psi(t_k)\right|_{t_{\ge 2}=0}$ is the $n=0$ term. By explicitly substituting the integral expression for the power series $A(z)$, we obtain
\begin{equation}
\left. \Psi(t_k)\right|_{t_{\ge 2}=0} = \int ds\, e^{-\alpha \frac{s^{p+1}}{p+1} } \frac{\alpha p}{2\pi}\int dz\, z^{p-1} e^{\alpha\, z^p (s-t_p)} % \cr
= \int ds\, e^{-\alpha \frac{s^{p+1}}{p+1} }\delta (s-t_p) 
%= e^{-\alpha \frac{t_p^{p+1}}{p+1} }\cr
%&
=1
\end{equation}
where the last equality follows from the fact that all higher times, amongst which is $t_p$, are set to zero. Thus, we have proven the initial condition. Therefore we have proven the identification \eqref{GeometricMainResult} between the matrix and the geometric open/closed generating functions.

\subsubsection*{Further Constraints}
Finally, let us remark that our Virasoro constraints $L_{-1}^{\op}$ and $L_0^{\op}$ correspond to the string equation and dilaton equation of \cite{Buryak:2018ypm} (after using charge conservation, or the dimension constraint).
Our Virasoro generators $L_n^{\op}$ extend these geometrically proven constraints to an infinite set.

\section{Conclusions}
\label{summary}
\label{conclusions}

Topological gravity coupled to topological matter is well-studied \cite{Witten:1989ig}. Introducing D-branes and applying concepts from holography  has allowed for a better understanding of these simple string theories, and has provided hands-on illustrations of profound concepts \cite{McGreevy:2003kb,Gaiotto:2003yb}.  Certainly, it has been very useful to underpin these achievements with the rigorous mathematical definition of topological gravity theories on Riemann surfaces with boundaries \cite{Pandharipande:2014qya}. This has stimulated progress in the identification of open/closed string correlators on Riemann surfaces with boundaries, and their relation to integrable systems \cite{Buryak:2018ypm}. This has in turn  allowed for a more rigorous understanding \cite{BCT1} of the  notion that boundaries can be replaced by closed string insertions  \cite{Polchinski:1995mt}.

Amongst the many insights that string theory has provided into topological gravity is the fact that the integration variables in the Kontsevich matrix model correspond to open strings. Indeed, in \cite{Maldacena:2004sn}, it was argued that the degrees of freedom in the Kontsevich matrix integral are  mesons made of open strings stretching between extended  and localized branes. In string theory, it is often useful to split a given set of D-branes into a heavy stack, and a single ``probe brane". In this work  we adopted such a probe brane analysis in treating the Kontsevich matrix model. We integrated out the open strings stretching between a large $N$ stack, and one extra D-brane. In doing so we have  naturally produced a matrix model realization of open/closed duality. The process of integrating out  generates a determinant operator insertion corresponding to the addition of a D-brane. From the integrable systems perspective this was argued in the case of  pure gravity in \cite{Alexandrov:2014gfa}.
It is important to mention that the insertion is distinct from the more familiar determinant operator in the Kazakov matrix model before taking the double scaling limit \cite{Maldacena:2004sn, Hashimoto:2005bf}. 

The hands-on treatment of the matrix model formulae allowed for a very precise treatment of the duality, the constraints the generating functions satisfy, and their relation to integrable systems. It is this relation to integrable systems that allowed us to precisely match our treatment to the rigorous algebro-geometric treatment of the $p$-spin intersection numbers on moduli spaces of Riemann surfaces with boundaries.
We note that both the extension of the closed string partition function that depends on all times of the KP integrable hierarchy \cite{BCT1} as well as the wave function that appears  in the integrable system \cite{Bertola:2014yka} pop out of our matrix model analysis spontaneously.

One direction for future research amongst many is to carefully match the geometric analysis with a first principle string field theory or conformal field theory derivation of the amplitudes on Riemann surfaces with boundaries.

\section*{Acknowledgments}
We thank our colleagues for creating a stimulating research environment, and Alexandr Buryak for patient explanations of the results in \cite{Buryak:2015eza}. SA would like to thank the \'Ecole Normale Sup\'erieure, Paris and the Universit\`a di Torino, Italy for their  hospitality during the completion of this work. 

\appendix

\section{Properties of the Fourier  Transform}
\label{FourierTransform}
\label{FT}
\label{LaplaceTransform}
\label{LT}

We exhibit some basic properties of the generalized Fourier transform that  prove useful in the analysis of the Virasoro generators. We recall the definition of the Fourier transform:
\begin{align}
\Phi[g(s)](z)& := \sqrt{\frac{\alpha\, p}{2 \pi}}\,  z^{\frac{p-1}{2}} \int ds\, e^{\alpha \left( \frac{z^{p+1}}{p+1} + z^p s\right)} \, g(s+z) \, .
\end{align}
We apply the transform to a function $g$ that is an $s$-derivative, $g(s) = \p_s f$:
\begin{align}
\label{GFTprop1}
\Phi\left[\frac{\p f}{\p s} \right] &= \sqrt{\frac{\alpha\, p}{2 \pi}}\,  z^{\frac{p-1}{2}} \int ds\, e^{\alpha \left( \frac{z^{p+1}}{p+1} + z^p s\right)} \, \frac{\p f(s+z)}{\p s}\cr
&= -\alpha\, z^p \Phi[f(s) ](z)\,,
\end{align}
where we have integrated by parts and used convergence properties of the functions under consideration. 

Secondly, we begin by calculating the $z$-derivative of the Fourier transform: 
\begin{align}
\frac{\p}{\p z} \Phi[f(s) ](z) = \frac{p-1}{2z} \Phi[f(s) ](z) + \sqrt{\frac{\alpha\, p}{2 \pi}}\,  z^{\frac{p-1}{2}}  \int ds\ \left( p\ \alpha\, z^{p-1} \, s + \alpha \, z^p + \frac{\p}{\p z} \right)f(s+z) \, .
\end{align}
The last two terms cancel against each other by first converting the $z$-derivative into an $s$-derivative and then using the  relation derived in equation \eqref{GFTprop1}. In the remaining term, we add and subtract $p\, \alpha\, z^p$ within the integrand, and obtain the inverse Fourier transform of the function $s\, f(s)$ on the right hand side. Rearranging the terms suitably, we finally obtain 
\begin{equation}
\Phi\left[ s\, f(s) \right](z) =\left( \frac{1}{\alpha\, p\, z^{p-1}} \frac{\p}{\p z}  - \frac{(p-1)}{2\, \alpha\, p\, z^p} + z \right) \Phi[ f(s)](z) \, .
\end{equation}
We summarize these results schematically:
\begin{align}
\label{GFTprop12}
-\alpha z^p &\longleftrightarrow \frac{\p}{\p s} \cr
\left( \frac{1}{\alpha\, p\, z^{p-1}} \frac{\p}{\p z}  - \frac{(p-1)}{2\, \alpha\, p\, z^p} + z \right) &\longleftrightarrow s \,.
\end{align}

\section{An Equivalent Virasoro Algebra}
\label{AltVirasoro}
\label{AlternativeVirasoro}
In this appendix, we provide the formulas that accompany the point of view that the (non-extended) closed string partition function should be strictly dependent on the couplings $t_{k \notin p \mathbb{Z}}$ only. In particular, this perspective gives rise to another form of the Virasoro algebra acting on the open/closed partition function dual to the pure closed partition function.

Recall that the extended Virasoro generators were obtained by conjugating the usual Virasoro generators by $e^{\xi}$ given in \eqref{xidefn}. Since this function involved all the times including the $t_{np}$, the extended Virasoro generators necessarily involved the times $t_{np}$ and associated derivatives. This is in contrast to the closed string partition function; defined via the generalized Kontsevich matrix model the partition function is necessarily independent of these times \cite{Kontsevich:1992ti, Itzykson:1992ya}. It is therefore also instructive to define a new wave potential $\widetilde{\tau}$ that is obtained by conjugation using an exponent $\widetilde{\xi}$ that does not involve these times $t_{np}$, and investigate how the closed-open Virasoro operators are related in this picture. This will  make contact with the existing literature \cite{Dijkgraaf:1991qh, Buryak:2015eza, Dijkgraaf:2018vnm}.

We begin by defining 
\begin{equation}
\label{tautildedefn}
\widetilde{\tau}(z,\Lambda) = e^{\widetilde{\xi}}\, \tau(\Lambda_z) \,,
\end{equation}
where 
\begin{align}
\label{xitildedefn}
\widetilde{\xi}  &= \log\det\left(1-\frac{z}{\Lambda}\right) -\frac{1}{p}\log \det \left(1-\frac{z^p}{\Lambda^p}\right) \cr
&= -\sum_{\substack {k=1 \\ k\ne 0\, \text{mod}\, p}}^{\infty} \frac{z^k}{k} \Tr \Lambda^{-k}  = \sum_{{\substack {k=1 \\ k\ne 0\, \text{mod}\, p}}}^{\infty} t_k\, z^k\,.
\end{align}
The modified statement of open-closed duality is the following compact equation:
\begin{equation}
\label{tautildedefn1}
\widetilde{\tau}(z,\Lambda) = \Phi[\widetilde{\tau}^{\op+\cl}(s)](z) \,,
\end{equation}
where $\Phi$ is the same generalized Fourier transform defined previously and the open/closed partition functions are different from the ones obtained earlier due to the extra factor that eliminated the $t_{np}$ times. In particular, we have: 
\begin{align}
\label{tauopenclosedtilde}
\widetilde{\tau}^{\op+\cl}(\Lambda, s) =  \frac{1}{D(\Lambda)} \int [dM] e^{-\alpha\frac{1}{(p+1)}\Tr \left [(M+\Lambda)^{p+1} \right]_{\ge 2} }\, {\mathcal O} \cdot e^{-\alpha\frac{1}{p+1} s^{p+1}}\,  \text{det}(M+\Lambda-s) \,.
\end{align}
The operator ${\mathcal O}$ is given by the differential operator $\det(\Lambda^p + \frac{1}{\alpha} \frac{\p}{\p s})^{-\frac{1}{p}}$ obtained by translating the $z^p$ operator into an operator acting on the open string variables using the results in appendix \ref{FT}. 

The open Virasoro generators one eventually obtains agree with those in the literature as we now show in brief.
We begin with the closed string Virasoro generators in equation \eqref{oldVirasoro}. The action of these generators on $\widetilde{\xi}$ gives us:
\begin{align}
L_n\,  e^{\widetilde{\xi}} &= \left( \alpha\, z^{np+p+1} + \frac{1}{p}\sum_{\substack {k=1 \\ k\ne 0\, \text{mod}\, p}}^{\infty}k t_k z^{np+k} + \frac{1}{2p} \sum_{\substack {k=1 \\ k\ne 0\, \text{mod}\, p}}^{np-1} z^{np} \right)e^{\widetilde{\xi}} \cr
& = \left( \alpha\, z^{np+p+1} + \frac{z^{np+1}}{p}\frac{\p}{\p z}+ \frac{n(p-1)}{2p} z^{np}  \right)e^{\widetilde{\xi}}
\end{align}
The main difference compared to the calculation in the main text is that the finite sum in the last term on the right hand side is over all integers from $1$ to $np-1$, but not including the integers that are $0$ mod $p$. This leads to a different coefficient for the $z^{np}$ term. The rest of the calculation proceeds as before and we obtain
\begin{align}
\label{Lclzmodified}
L_{n}  \widetilde{\tau}(z, \Lambda)   
& = \left( \alpha z^{np+p+1}   + \frac{1}{p}  z^{np+1} \frac{\p}{\p z} + \frac{ n(p-1)}{2p} \, z^{np} \right) \widetilde{\tau}(z,\Lambda) \cr
&= \alpha\, z^{(n+1)p}\left(z+\frac{1}{\alpha\, p\, z^{p-1}}\frac{\p}{\p z}-\frac{(p-1)}{2\alpha p\, z^p}\right)\widetilde{\tau}(z, \Lambda)+\frac{(n+1)(p-1)}{2p}\, z^{np}\widetilde{\tau}(z, \Lambda) \,. \cr
\end{align}
For the generator $L_0$, we similarly find 
\begin{align}
\label{L0zmodified}
L_{0}  \widetilde{\tau}(z, \Lambda)   
& = \left( \alpha z^{p+1}   + \frac{1}{p}  z\, \frac{\p}{\p z} \right) \widetilde{\tau}(z,\Lambda)\cr
& =\alpha\, z^p \left( z  + \frac{1}{\alpha\, p z^{p-1}}  \, \frac{\p}{\p z}-\frac{(p-1)}{2\alpha p\, z^p} \right) \widetilde{\tau}(z,\Lambda) + \frac{p-1}{2p}\, \widetilde{\tau}(z,\Lambda) \, .
\cr
\end{align}
Lastly, for the $L_{-1}$ operator, we find the same result as before: 
\begin{align}
\label{Lminus1zmodified}
L_{-1}  \widetilde{\tau}(z, \Lambda)   
& =  \alpha\left(z + \frac{1}{\alpha\, p\, z^{p-1}} \frac{\p}{\p z}  -\frac{p-1}{2\alpha\, p\, z^p } \right)  \widetilde{\tau}(z,\Lambda) \,.
\end{align}
By utilizing the properties of the generalized Fourier transform  \eqref{GFTprop12}, we finally obtain the expression for the open string Virasoro generators acting on the modified open/closed partition function:
\begin{align}
% L_{-1}^{\op}(s) &= -\alpha\, s\cr
% L_0^{\op}(s) &= s\, \frac{\p}{\p s} - \frac{p-1}{2p} = \left(s\frac{\p}{\p s} + \frac{p+1}{2p} \right)\cr
L_n^{\op}(s) &=  (-\alpha)^{-n} \left( \frac{\p^{n+1}}{\p s^{n+1}} \cdot s -  \frac{(p-1)}{2p}(n+1) \frac{\p^n}{\p s^n} \right)\cr
& =  (-\alpha)^{-n} \left( s\frac{\p^{n+1}}{\p s^{n+1}} +  \frac{(p+1)}{2p}(n+1) \frac{\p^n}{\p s^n} \right)
\end{align}
For $p=2$, this agrees with the Virasoro generators obtained in \cite{Buryak:2015eza, Dijkgraaf:2018vnm}.

\section{Another Pure Duality}
\label{puregravity}
\label{PureAlternative}
\label{alternativepuregravity}
In this appendix, we review an alternative expression of open/closed duality derived in the mathematical literature for the $p=2$ case. The derivation has features in common with  our derivation in section \ref{integratingout}, but it also differs in important aspects.  The rigorous mathematical treatment of this version of the duality is recent \cite{Buryak:2015eza,BCT1,Buryak:2018ypm}. The string theoretic conceptual background is older \cite{McGreevy:2003kb,Gaiotto:2003yb,Maldacena:2004sn}. 
 We refer to the literature for much of the background, and merely draw attention to the alternative derivation, on the one hand, and on the other hand, the consequent differences in the final expression of the duality.

The starting point in \cite{Buryak:2015eza} is also the Kontsevich matrix model. However the Harish-Chandra-Itzykson-Zuber formula is applied to the matrix integral with an exponent quadratic in the matrix variable of integration.\footnote{This approach is only directly applicable to the case of pure gravity, namely $p=2$.} This differs from the road we pursued in the bulk where we performed the Harish-Chandra-Itzykson-Zyber integration for an exponent linear in the matrix variable of integration for all $p$. This results in a surprisingly different formula for the open/closed string partition function. Firstly, the closed string partition function naturally becomes independent of the even times, after combining with the determinants, and the combination equals a formal Fourier transform of the open/closed partition function -- we refer to \cite{Buryak:2015eza} for the detailed formulas sketched in this appendix --:
\begin{equation}
\label{BTopenclosed}
\sqrt{ \frac{\det(1- \frac{z}{\Lambda})}{\det(1+\frac{z}{\Lambda})}}\, \tau(\Lambda_z)  =
 \Phi_s^{form} [ \tilde{\tau}^{\op+\cl}] (\Lambda,z) \, .
\end{equation}
The formal Fourier transform $\Phi^{form}_s$ is a close analogue of our transform $\Phi$ \cite{Buryak:2015eza}.
Note that the left hand side matches the function $\widetilde{\tau}$ we defined in equation \eqref{tautildedefn} when $p=2$. The modified open string partition function, though, takes the form  \cite{Buryak:2015eza}:
\begin{equation}
\label{tauopenclosedBT}
\widetilde{\tau}^{\op+\cl}(\Lambda,z) =
%\frac{1}{D(\Lambda)} %\sqrt{\frac{z}{2 \pi}} \int
%_{\mathbb{R}} 
%e^{\frac{i}{6} s^3 - \frac{1}{2} s^2 z} 
\int d \mu(s) \int %_{H}
d \mu (M)
\det \left(\frac{\Lambda+\sqrt{\Lambda^2-z^2}-M+s-z}{\Lambda+\sqrt{\Lambda^2-z^2}-M-s+z} \right) \, . 
\end{equation}
We observe an insertion of a ratio of determinants, corresponding to both a brane  and an anti-brane insertion. It would be interesting to see whether the formula  \eqref{tauopenclosedBT} allows for a generalization to larger $p$. Moreover, given the equality of the closed string sides in \eqref{tautildedefn1} and \eqref{BTopenclosed}, it would be important to understand the relation between the brane/anti-brane operator in equation \eqref{tauopenclosedBT} and the   insertion we obtained in expression \eqref{tauopenclosedtilde}. While in the case of pure gravity, the relation can be established by retracing all steps in both lines of reasoning, a more direct physical understanding would be welcome.

\bibliographystyle{JHEP}

\end{document}